\documentclass[twocolumn,showpacs,preprintnumbers,amsmath,amssymb,prb]{revtex4}
\usepackage{graphicx}% Include figure files
\usepackage{dcolumn}% Align table columns on decimal point
\usepackage{bm}% bold math

\begin{document}

\preprint{APS/123-QED}

\title{Ferromagnetism induced in anisotropic stacked kagome-lattice antiferromagnet Cs$_2$Cu$_3$CeF$_{12}$}% Force line breaks with \\

\author{Taiki Amemiya$^1$}
%\email{amemiya@lee.phys.titech.ac.jp.}
\author{Izumi Umegaki$^1$}
%\email{umegaki@lee.phys.titech.ac.jp.}
\author{Hidekazu Tanaka$^1$} 
%\email{tanaka@lee.phys.titech.ac.jp.}
\author{Toshio Ono$^2$}
%\email{o-toshio@lee.phys.titech.ac.jp.}
\author{Akira Matsuo$^3$}
%\email{a-matsuo@issp.u-tokyo.ac.jp}
\author{Koichi Kindo$^3$}
%\email{kindo@issp.u-tokyo.ac.jp}
\affiliation{
$^1$Department of Physics, Tokyo Institute of Technology, Meguro-ku, Tokyo 152-8551, Japan\\
$^{2}$Department of Physics, Osaka Prefecture University, Sakai, Osaka 599-8531, Japan\\
$^3$Institute for Solid State Physics, University of Tokyo, Kashiwa, Chiba 277-8581, Japan
}

\date{\today}% It is always \today, today,

\begin{abstract}
The magnetic properties of Cs$_2$Cu$_3$CeF$_{12}$ were investigated through magnetization and specific heat measurements. Cs$_2$Cu$_3$CeF$_{12}$ is composed of a buckled kagome lattice of Cu$^{2+}$, which is stacked along the $b$ axis. The exchange network in the buckled kagome lattice is strongly anisotropic. Consequently, Cs$_2$Cu$_3$CeF$_{12}$ can be divided into two subsystems: alternating Heisenberg chains with strong antiferromagnetic exchange interactions and dangling spins. The dangling spins couple with one another via effective exchange interactions, which are mediated by chain spins. The dangling spins are further divided into two subsystems, DS1 and DS2. The dangling spins in DS1 undergo three-dimensional ferromagnetic ordering at 3.14\,K, while those in DS2 remain paramagnetic down to 0.35\,K. The effective interaction between the DS1 spins is approximately expressed by the ferromagnetic $XXZ$ model with the $z$ direction parallel to the crystallographic $c$ axis. A magnetic phase diagram for $H\,{\parallel}\,c$ was obtained and was analyzed within the framework of the molecular field approximation. With increasing magnetic field, the dangling spins are polarized and the magnetization curve exhibits a wide plateau at one-third of the saturation magnetization.
\end{abstract}

%Use showkeys class option if keyword
\maketitle

%--------------------------------------------------------------------------------------------------------------------

\section{Introduction}
Quantum kagome-lattice antiferromagnets (KLAFs) are at the frontier of research on magnetism.\cite{ML} For the last two decades, the natures of the ground state and excitations for spin-$1/2$ Heisenberg KLAFs have been actively studied theoretically.\cite{Zeng1,Sachdev,Elstner,Nakamura,Lecheminant,Waldtmann,Syromyatnikov,Jiang} It was demonstrated that the interplay between spin frustration and strong quantum fluctuation leads to a disordered ground state. However, its nature is still unresolved. A valence-bond solid (VBS) described by a periodic arrangement of singlets,\cite{Singh2,Nikolic,Budnik,Yang} a resonating valence-bond (RVB) state given by a linear combination of various configurations of singlets\,\cite{Mambrini,Hastings} and a gapless critical spin liquid\,\cite{Ryu,Hermele} have been proposed as the ground state of spin-$1/2$ Heisenberg KLAFs. %The existence of the triplet gap is still under debate.

Experimental studies on spin-$1/2$ kagome antiferromagnets have been performed on several materials. The hexagonal A$_2$Cu$_3$MF$_{12}$ family with ${\rm A}\,{=}\,{\rm Cs}$ and Rb and ${\rm M}\,{=}\,{\rm Zr}$, Hf and Sn is a promising candidate of quantum KLAFs because fairly large high-quality single crystals can be obtained,\cite{Mueller,Yamabe,Morita,Ono,Matan} which enable microscopic studies to be carried out. A gapped singlet ground state composed of a pinwheel VBS was observed in Rb$_2$Cu$_3$SnF$_{12}$.\cite{Morita,Ono,Matan} A magnetic susceptibility close to the theoretical susceptibility was also observed in the paramagnetic phase of Cs$_2$Cu$_3$SnF$_{12}$.\cite{Ono}
 
\begin{figure}[t]
\includegraphics[width=7.5cm, clip]{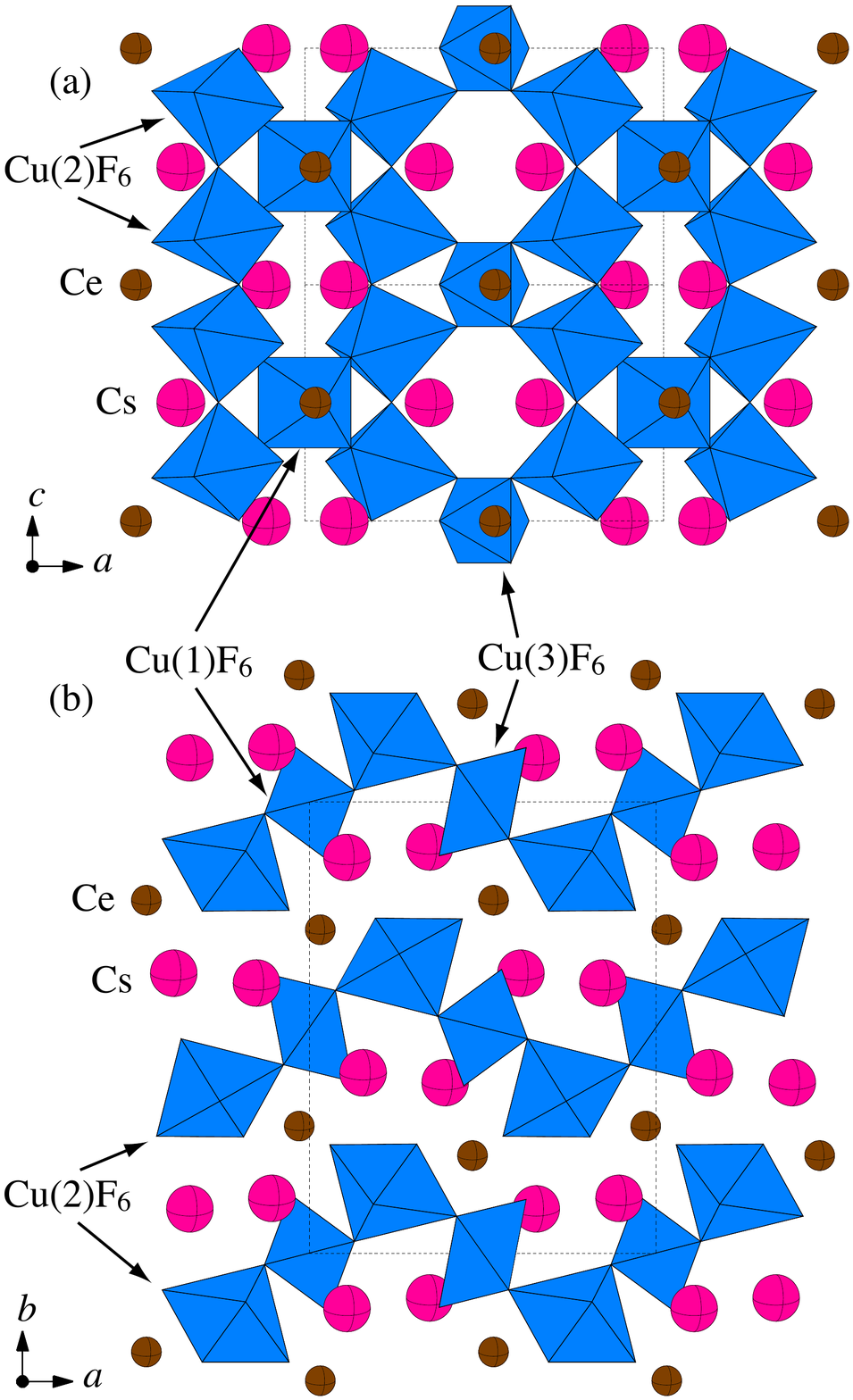}
\caption{(Color online) Crystal structure of Cs$_2$Cu$_3$CeF$_{12}$ viewed (a) along the $b$ axis and (b) along the $c$ axis. Shaded octahedra represent CuF$_6$ octahedra. Dotted lines denote the chemical unit cell. Cu$^{2+}$ ions form a buckled kagome lattice in the $ac$ plane.}
 \label{fig:structure}
 \end{figure}

This paper deals with the unusual magnetic ordering observed in Cs$_2$Cu$_3$CeF$_{12}$, which is composed of buckled kagome layers of Cu$^{2+}$. Cs$_2$Cu$_3$CeF$_{12}$ was synthesized in the process of developing the above-mentioned A$_2$Cu$_3$MF$_{12}$ kagome family. The crystal structure of Cs$_2$Cu$_3$CeF$_{12}$ is orthorhombic, $Pnnm$.~\cite{Amemiya} Figures~\ref{fig:structure}(a) and (b) show the crystal structures of Cs$_2$Cu$_3$CeF$_{12}$ viewed along the $b$ axis and $c$ axis, respectively. All Cu$^{2+}$ ions are surrounded octahedrally by six F$^-$ ions. Because of the Jahn-Teller effect, the CuF$_6$ octahedra centered at Cu(2) and Cu(3) are elongated along one of the principal axes, while those centered at Cu(1) are compressed. Consequently, the orbital ground state for Cu(2) and Cu(3) is $d(x^2\,{-}\,y^2)$, while that for Cu(1) is $d(3z^2\,{-}\,r^2)$. Cu$^{2+}$ ions form a spatially anisotropic kagome lattice in the $ac$ plane, which is buckled and has the appearance of a staircase, as shown in Fig.~\ref{fig:structure}(b). 

%\vspace{-0.5cm}

\begin{figure}[t]
\includegraphics[width=8.0 cm, clip]{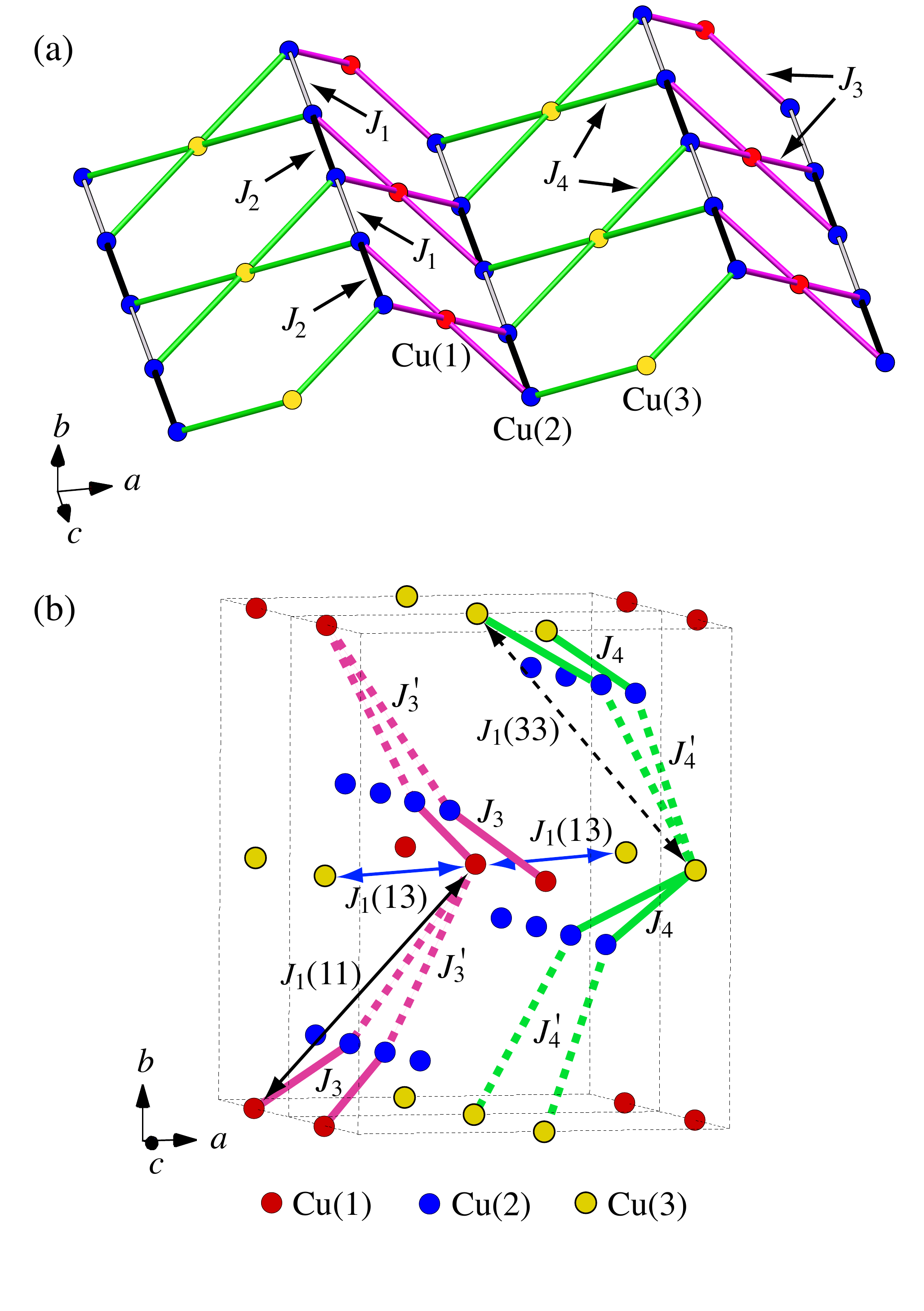}
\caption{(Color online) (a) Exchange network on the buckled kagome lattice parallel to the $ac$ plane in Cs$_2$Cu$_3$CeF$_{12}$. (b) Interlayer exchange interactions $J^{\prime}_3$ and $J^{\prime}_4$ (dashed lines), and effective exchange interactions between dangling spins ${\cal J}_1(11)$, ${\cal J}_1(33)$ and ${\cal J}_1(13)$ (double-headed arrows). Dotted lines denote the chemical unit cell. }
 \label{fig:exchange}
 \end{figure}

Figure~\ref{fig:exchange}(a) shows the exchange network on the buckled kagome lattice. The spatial anisotropy in the kagome lattice produces inequivalent spin sites. There are four types of neighboring exchange interaction, $J_1\,{-}\,J_4$. Because the hole orbitals $d(x^2\,{-}\,y^2)$ of Cu(2) ions are linked through $p$ orbitals of F$^{-}$ ions along the $c$ axis and the Cu(2)${-}$F${-}$Cu(2) bond angle is approximately $149^{\circ}$, the exchange interactions $J_1$ and $J_2$ between Cu(2) ions are antiferromagnetic and strong, $J_1/k_{\rm B}\,{\simeq}\,J_2/k_{\rm B}\,{\sim}\,300$~K.\cite{Morita,Ono,Amemiya} Because the Cu(2)${-}$Cu(2) distance alternates slightly along the $c$ axis, the exchange interactions $J_1$ and $J_2$ are alternating. The magnitude of $J_2$ has been estimated to be approximately $J_2/J_1\,{\simeq}\,0.9$.\cite{Amemiya} The exchange interactions $J_3$ and $J_4$ are much smaller than $J_1$ and $J_2$ because the hole orbitals of Cu(1) and Cu(3) ions are not directly linked to those of Cu(2) ions through $p$ orbitals of F$^{-}$ ions. Therefore, Cs$_2$Cu$_3$CeF$_{12}$ is described as a strongly spatially anisotropic spin-$1/2$ Heisenberg-like KLAF, which can be divided into two subsystems: weakly alternating Heisenberg chains with strong antiferromagnetic exchange interactions and dangling spins. Because the dangling spins undergo three-dimensional (3D) ordering as shown later, the interlayer interactions $J_3^{\prime}$ and $J_4^{\prime}$ are necessary.
 
The ground state for spatially anisotropic KLAFs has been discussed theoretically.\cite{Wang,Ya,Sindzingre,Stoudenmire,Schnyder} When the exchange network becomes anisotropic, the spin frustration may be reduced. However, such a spatially anisotropic model will be useful for obtaining a deep understanding of the quantum effect in KLAFs. Some unusual ground states have been predicted for extremely anisotropic cases.\cite{Ya,Stoudenmire,Schnyder} It was also found in the spin-$1/2$ triangular-lattice antiferromagnet Cs$_2$CuBr$_4$ that spatial anisotropy in the exchange interactions enhances the frustration effect and leads to many quantum states in magnetic fields including magnetization plateaus at one-third and two-thirds of the saturation magnetization.\cite{Ono1,Ono2,Miyahara,Fortune}  

In our previous paper, \cite{Amemiya} we briefly reported the crystal structure and magnetic properties of Cs$_2$Cu$_3$CeF$_{12}$. In this paper, we report the results of magnetization and specific heat measurements in detail and discuss the 3D ordering of the dangling spins. As shown below, it was found that the dangling spins are further divided into two subsystems, one of which undergoes 3D ordering at 3.1\,K, while the other remains paramagnetic down to 0.35 K. The effective exchange interaction that gives rise to the 3D ordering is described by the spin-$1/2$ ferromagnetic $XXZ$ model with the $z$ direction parallel to the $c$ axis, and the 3D ordering of dangling spins for a magnetic field parallel to the $c$ axis can be understood as the Bose-Einstein condensation (BEC) of lattice bosons, as discussed by Matsubara and Matsuda.~\cite{Matsubara} 

 This paper is organized as follows: the experimental procedure is given in section II. The results of magnetization and specific heat measurements and their analyses are shown in section III. In section IV, we discuss the ordering of dangling spins in a magnetic field. Section V is devoted to a conclusion.

%--------------------------------------------------------------------------------------------------------------------

\section{Experiments}
Cs$_2$Cu$_3$CeF$_{12}$ single crystals were grown from a melt of $\mathrm{CsF}$, $\mathrm{CuF_2}$ and $\mathrm{CeF_4}$. The materials were dehydrated by heating in vacuum at $60\,{-}\,$100$^{\circ}$C and were packed into a Pt tube in the ratio of $3\,{:}\,3\,{:}\,2$. One end of the Pt tube was welded and the other end was tightly folded with pliers. The Pt tube was placed about $10$\,cm from the center of a horizontal furnace to induce a temperature gradient in the tube. The temperature of the furnace was increased to 750$^{\circ}$C in 15 h then lowered from 750 to 500$^{\circ}$C over four days. Transparent light-blue crystals having a long platelet-like shape with a wide $(1,0,0)$ plane were obtained. From X-ray diffraction measurements, the long direction was found to be parallel to the $c$ axis.

Magnetization was measured in the temperature range of $1.8\,{-}\,400$ K using a SQUID magnetometer (Quantum Design MPMS XL). The specific heat was measured by the relaxation method in the temperature range of $0.35\,{-}\,200$ K using a Physical Property Measurement System (Quantum Design PPMS). High-field magnetization measurement in magnetic fields of up to 54 T was performed using an induction method with a multilayer pulse magnet at the Institute for Solid State Physics, University of Tokyo.
%--------------------------------------------------------------------------------------------------------------------

\section{Results and analyses}
\begin{figure}[htbp]
\vspace{-2cm}
\begin{center}
\hspace{-0.2cm}\includegraphics[width=8.5cm, clip]{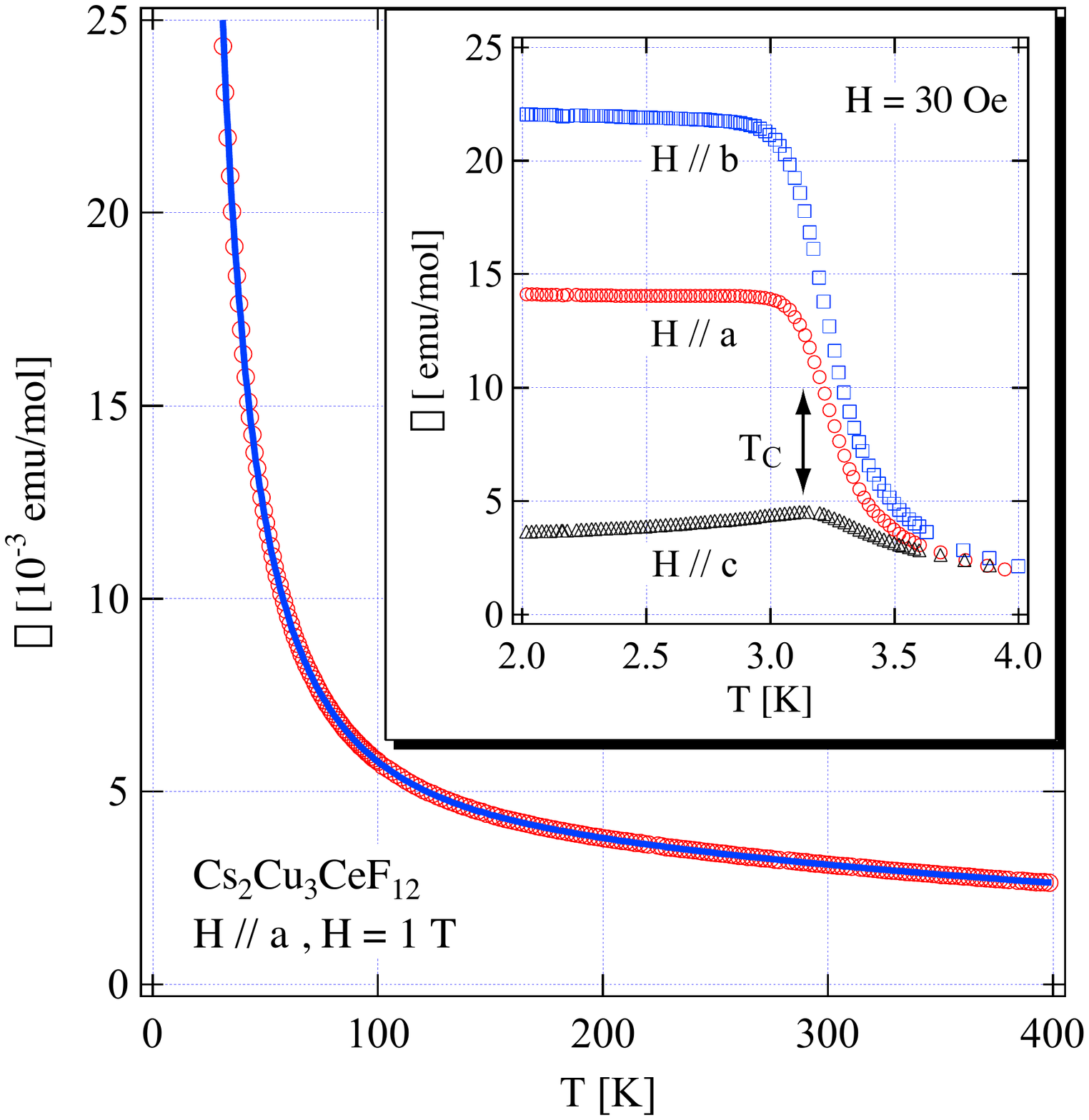}
\end{center}
\vspace{-2.5cm}
\caption{(Color online) Temperature dependences of magnetic susceptibility in Cs$_2$Cu$_3$CeF$_{12}$ measured at 1.0 T for $H\,{\parallel}\,a$. Solid line is the fit given by eqs.~(\ref{eq:sus1})$-$(\ref{eq:sus3}) with the exchange parameters and $g$ factors given in the text. The inset shows the low-temperature magnetizations measured at 30 Oe for $H\,{\parallel}\,a$, $b$ and $c$. The double-headed arrow indicates the Curie temperature $T_{\rm C}\,{=}\,3.14$ K.}
 \label{fig:sus}
 \end{figure}

In Fig.~\ref{fig:sus}, we show the temperature dependence of magnetic susceptibility in Cs$_2$Cu$_3$CeF$_{12}$ measured at 1.0 T for $H\,{\parallel}\,a$. The magnetic susceptibility above 200 K obeys the Curie-Weiss law with the Weiss constant ${\Theta}\,{\simeq}-250$ K, which is clearly seen in the inverse susceptibility vs temperature shown in Fig.~2 of our previous paper.\cite{Amemiya} This large Weiss constant arises from the spin chains with strong antiferromagnetic exchange interactions $J_1$ and $J_2$. The rapid increase in the magnetic susceptibility below 100 K is attributed to the paramagnetic susceptibility of the dangling spins, which are weakly coupled to the chain spins through $J_3$ and $J_4$ and also through interlayer exchange interactions $J^{\prime}_3$ and $J^{\prime}_4$ as shown in Fig.~\ref{fig:exchange}(b). The interlayer interactions are essential for discussing the magnetic ordering of the dangling spins. The analysis of the high-temperature susceptibility is given below.

The inset of Fig.~\ref{fig:sus} shows the low-temperature magnetizations measured at $H\,{=}\,30$ Oe for $H\,{\parallel}\,a$, $b$ and $c$. The magnetization exhibits a sharp increase between 3.5 and 3~K and becomes almost constant for $H\,{\parallel}\,a$ and $b$, while for $H\,{\parallel}\,c$, the magnetization exhibits cusplike peak at $T_{\rm C}\,{\simeq}\,3.2$ K. This indicates the occurrence of ferromagnetic ordering at $T_{\rm C}$.\cite{Amemiya} The Curie temperature was determined as $T_{\rm C}\,{=}\,3.14$ K by the present specific heat measurements. The details of the low-temperature magnetization in various magnetic fields are shown later.
%In Fig. \ref{fig:Neel}, we also show the low-temperature specific heat measured at zero field. A $\lambda$-like anomaly, indicative of magnetic ordering, is observed at $T_{\rm C}\,{=}\,3.14$ K. For $H\parallel a$, the cusp anomaly smears rapidly with increasing magnetic field, as shown in Fig.~\ref{fig:heat_a1}. This specific heat behavior in magnetic field is characteristic of the ferromagnetic ordering.
 
%\begin{figure}[htbp]
%\begin{center}
%\includegraphics[width=8.5cm, clip]{Neel.eps}
%\end{center}
%\caption{(Color online) Low-temperature magnetization measured at 30 Oe for $H\,{\parallel}\,a$ and low-temperature specific heat measured at zero field. Vertical arrows indicate the ordering temperature $T_{\rm C}$ determined from the specific heat measurement. }
%\label{fig:Neel}
%\end{figure} 

\begin{figure}[htbp]
\vspace{-2cm}
\begin{center}
\includegraphics[width=8.5cm, clip]{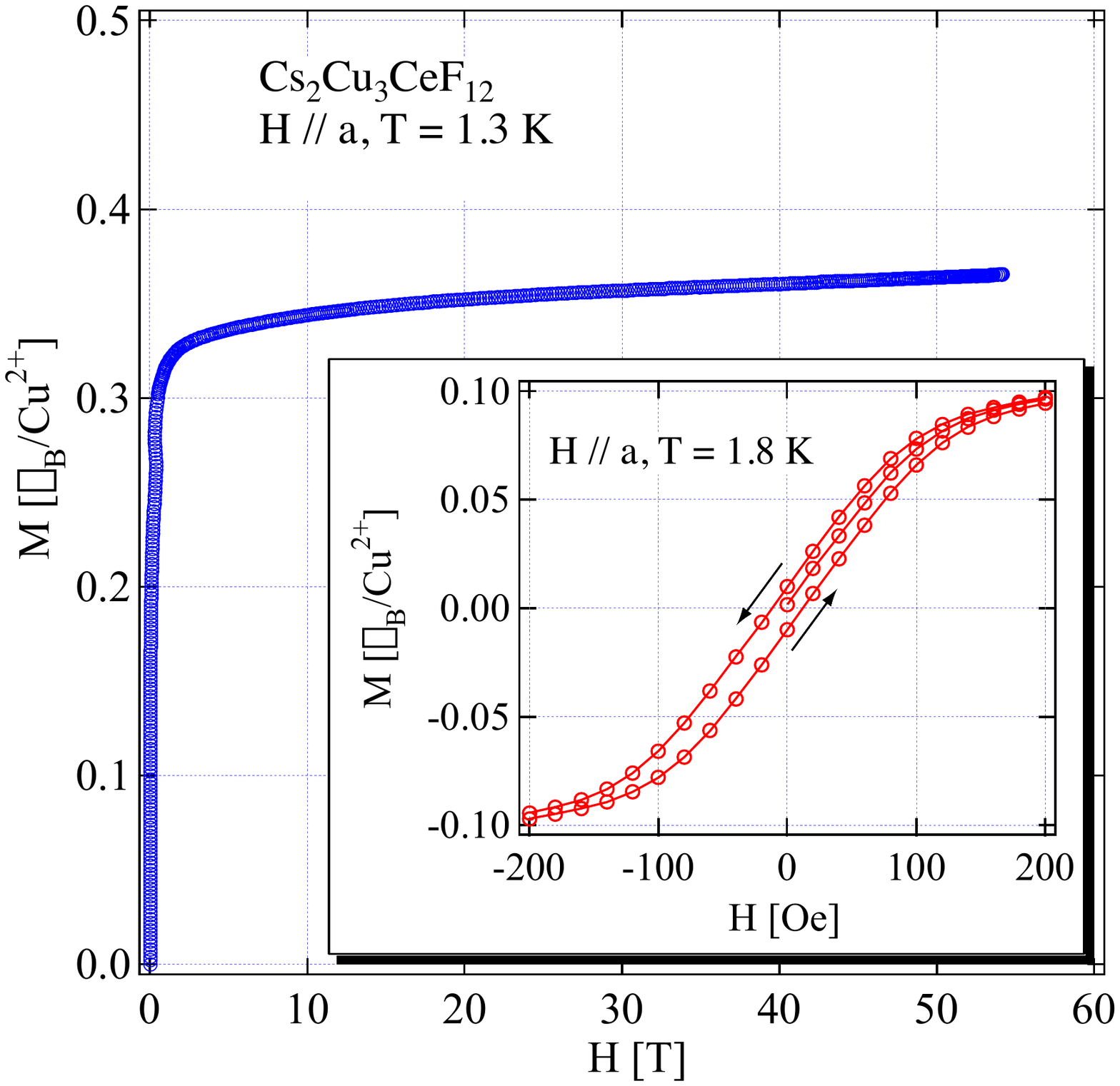}
\end{center}
\vspace{-3cm}
\caption{(Color online) Magnetization curve for $H\,{\parallel}\,a$ measured at 1.3 K in pulsed high magnetic field. The inset shows the hysteresis of the magnetization curve for $H\,{\parallel}\,a$ measured at 1.8 K using a SQUID magnetometer.}
\label{fig:MH}
\end{figure}

Figure~\ref{fig:MH} shows the magnetization curve measured in a pulsed high magnetic field at 1.3 K for $H\,{\parallel}\,a$. The data were collected during the downsweeping of the magnetic field. The absolute value of magnetization was corrected using the magnetization measured at 1.8 K in a static magnetic field. The correction was made by adding a constant value. %of 0.13 ${\mu}_{\rm B}/{\rm Cu}^{2+}$. 
Because the magnetization near $H\,{\simeq}\,0$ increases very rapidly with increasing magnetic field, $dM/dH$ exhibits a $\delta$-function-like anomaly near $H\,{\simeq}\,0$. This leads to an error of a constant value in the magnetization. 

The magnetization for $H\,{\parallel}\,a$ exhibits a plateau at one-third of the saturation magnetization. We also confirmed that the magnetization displays the $1/3$-plateau for $H\,{\parallel}\,b$ and $c$, using a SQUID magnetometer. Because the numbers of chain spins and dangling spins are in the ratio of $2\,{:}\,1$, we can deduce that this magnetization plateau arises from the full polarization of the weakly coupled dangling spins. It is considered that in the $1/3$-plateau state, the chain spins are in a gapped singlet state. The $1/3$-plateau state is robust and continues to exist below 54 T. From these observations, we can conclude that the magnetic phase transition at $T_{\rm C}\,{=}\,3.14$ K arises from the ferromagnetic long-range order of the dangling spins. The origin of the ferromagnetic exchange interaction between dangling spins is discussed later.

\begin{figure}[htbp]
\vspace{-2cm}
\begin{center}
\includegraphics[width=8.5cm, clip]{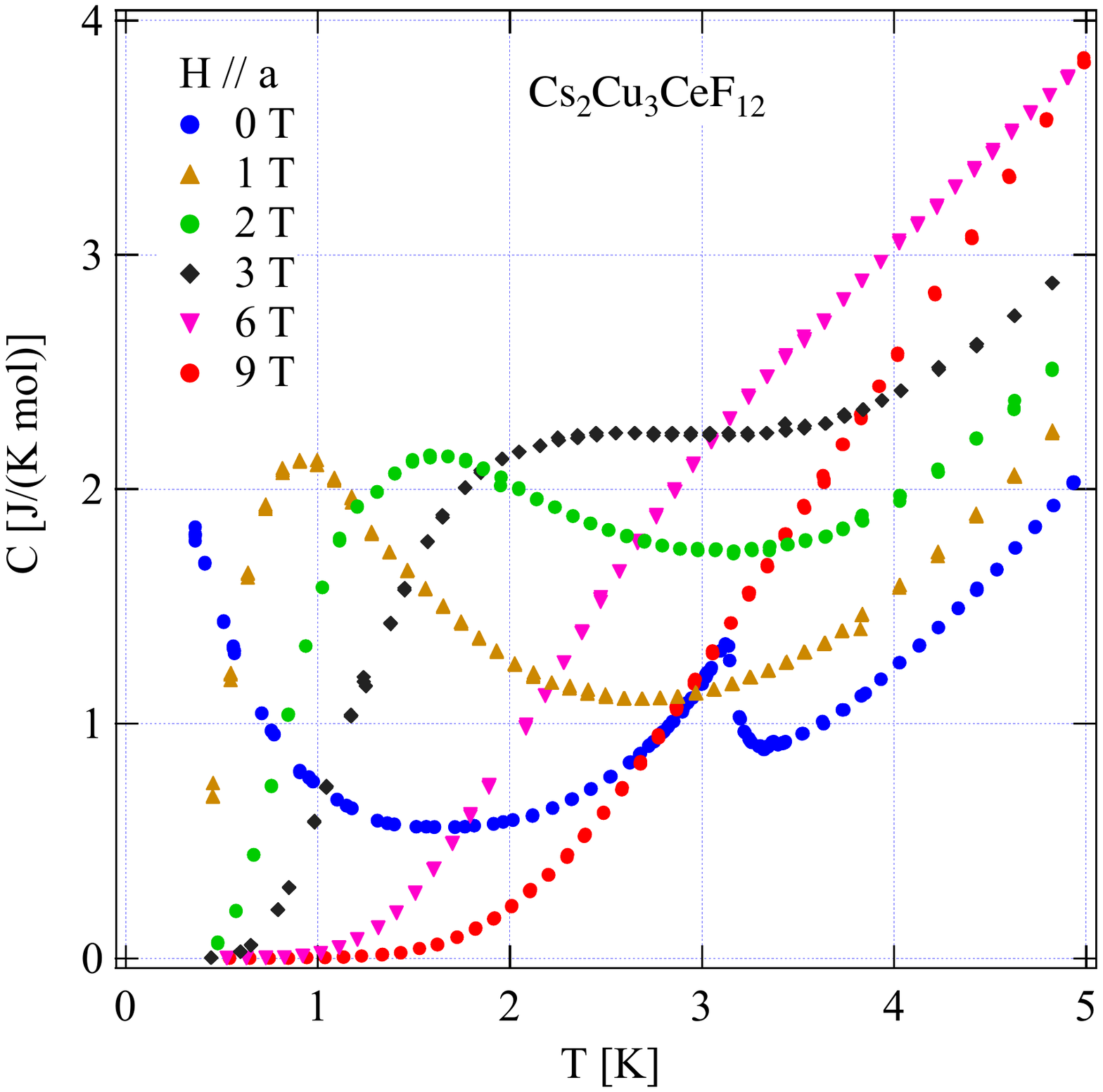}
\end{center}
\vspace{-3cm}
\caption{(Color online) Specific heat measured at zero magnetic field and at various large magnetic fields for $H\,{\parallel}\,a$. }
\label{fig:heat_a2}
\end{figure}

\begin{figure}[htbp]
\vspace{-2cm}
\begin{center}
\includegraphics[width=8.5cm, clip]{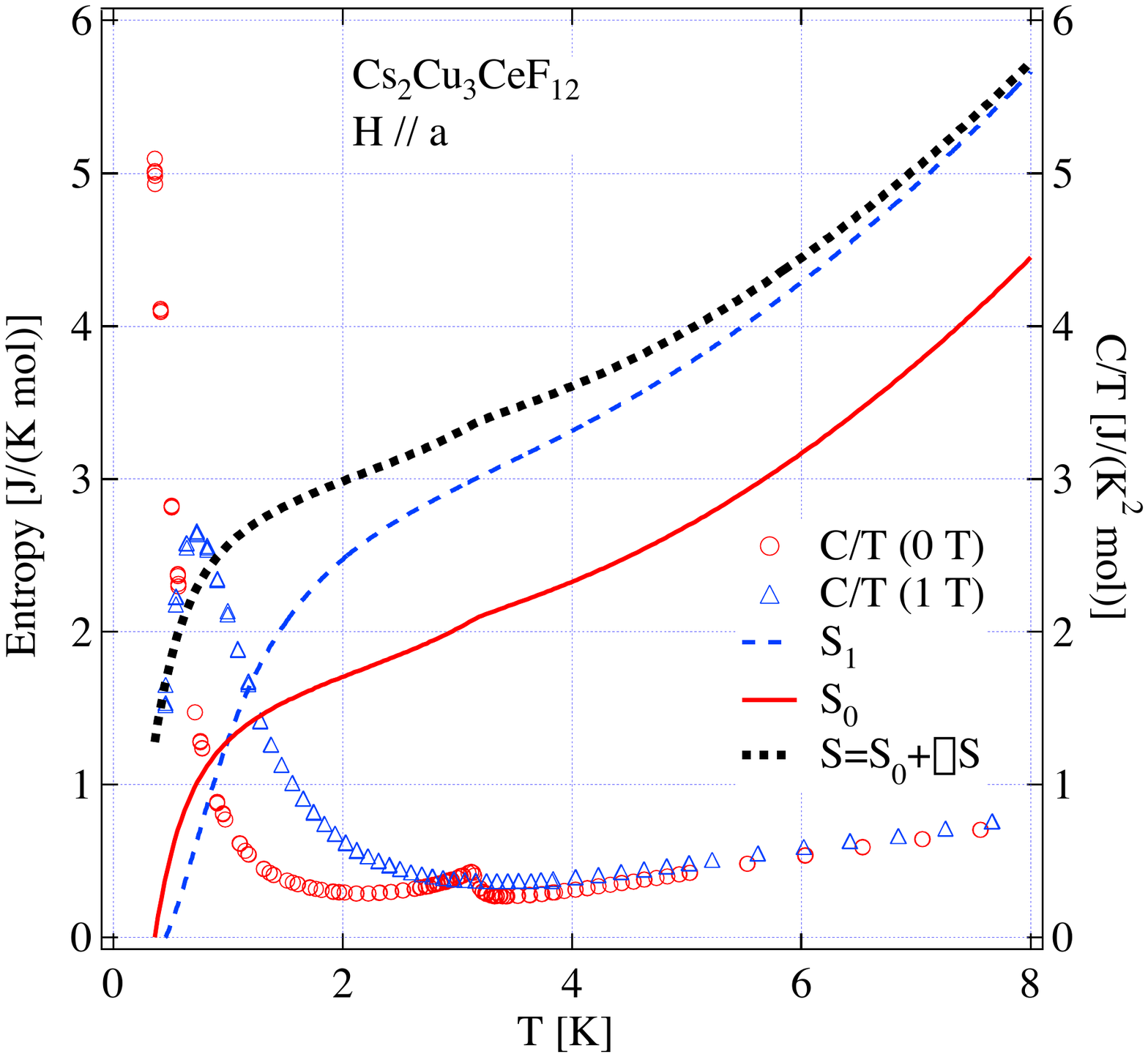}
\end{center}
\vspace{-3cm}
\caption{(Color online) Entropy and $C/T$ vs temperature measured at $H\,{=}\,0$ and 1 T for $H\,{\parallel}\,a$. Thick dotted line indicates the estimated entropy for $H\,{=}\,0$. For the definitions of $S_0$ and $S_1$ shown by thin solid lines, see text.}
\label{fig:entropy}
\end{figure}

Figure~\ref{fig:heat_a2} shows the total specific heat measured at zero magnetic field and at various large magnetic fields for $H\,{\parallel}\,a$. With decreasing temperature, the specific heat for $H\,{=}\,0$ has a minimum at 1.7 K after exhibiting the $\lambda$-anomaly at $T_{\rm C}\,{=}\,3.14$ K then increases rapidly with a tendency to diverge. At $H\,{=}\,1$\,T, the specific heat exhibits a rounded peak at $T_{\rm max}\,{\simeq}\,1$ K. The temperature $T_{\rm max}$ giving the rounded peak increases with increasing external magnetic field. This behavior is characteristic of the specific heat for nearly isolated spins. From these observations, we can deduce that part of the dangling spins (DS1) take part in the magnetic ordering at $T_{\rm C}$ and the remainder of the dangling spins (DS2) are still paramagnetic below $T_{\rm C}$. 
 
To estimate the fraction of the paramagnetic spins, we evaluate the low-temperature entropy for $H\,{=}\,0$. Figure~\ref{fig:entropy} shows $C/T$ vs $T$ for $H\,{=}\,0$ and 1 T. For zero magnetic field, $C/T$ increases rapidly below 1 K, and thus, the large entropy remains below the lowest experimental temperature of 0.36 K. For $H\,{=}\,1$ T, $C/T$ exhibits a rounded maximum at 0.7 K and decreases rapidly with decreasing temperature, which shows that little entropy remains below the lowest experimental temperature of 0.45 K. Thus, total entropy for $H\,{=}\,1$ T approximates
\begin{eqnarray}
\label{eq:entropy1}
S_1(T)=\int_{0.45\,{\rm K}}^T \frac{C(1\,{\rm T})}{T}\,dt.
\end{eqnarray}
We define $S_0(T)$ for $H\,{=}\,0$ as
\begin{eqnarray}
\label{eq:entropy0}
S_0(T)=\int_{0.36\,{\rm K}}^T \frac{C(0\,{\rm T})}{T}\,dt.
\end{eqnarray}
Thin solid and dashed lines in Fig.~\ref{fig:entropy} denote $S_0(T)$ and $S_1(T)$, respectively. The difference between $S_1(T)$ and $S_0(T)$ becomes constant at high temperatures, which is given by ${\Delta}S\,{\simeq}\,1.28$ J/(K\,mol). The ${\Delta}S$ should be close to the entropy below 0.36 K for $H\,{=}\,0$. Thus, the total entropy for $H\,{=}\,0$ is estimated as $S(T)\,{=}\,S_0(T)\,{+}\,{\Delta}S$, which is shown by thick dotted line in Fig.~\ref{fig:entropy}. The entropy $S_{\rm DS2}$ of the paramagnetic DS2 spins for $T\,{\rightarrow}\,\infty$ will be roughly given by $S(T\,{=}\,2\,{\rm K})$, because below 2 K, the lattice contribution to the specific heat is negligible and the entropy of the ordered DS1 spins and the chain spins is much smaller than $S_{\rm DS2}$. Therefore, we can estimate the entropy of the paramagnetic DS2 spins as $S_{\rm DS2}\,{\simeq}\,3.0\,{\rm J/(K\,mol)}\,{\simeq}\,(1/2)R\ln 2$. Because one molar Cs$_2$Cu$_3$CeF$_{12}$ contains three molar Cu$^{2+}$ ions, this result means that one-half of the dangling spins are the paramagnetic DS2 spins. Because the Cu(1) and Cu(3) sites are magnetically different, it is natural to consider that the spins on either the Cu(1) site or Cu(3) site should be DS1 spins. %As discussed later, the spins on Cu(1) and Cu(3) should be DS1 and DS2 spins, respectively.

Next we reevaluated the individual exchange parameters from the magnetic susceptibility data shown in Fig.~\ref{fig:sus}. Applying the mean field approximation to the exchange interactions between chain spins and dangling spins, we express the magnetic susceptibilities of chain spins and dangling spins as
\begin{eqnarray}
\label{eq:sus1}
&&{\chi}_{\rm ch}={\chi}_{\rm ch}^0\frac{1-\displaystyle\frac{6}{Ng_{\rm ch}g_{\rm d}{\mu}_{\rm B}^2}\left({\tilde J}_3{\chi}_{\rm d1}^0+{\tilde J}_4{\chi}_{\rm d3}^0\right)}{1-\displaystyle\frac{36{\chi}_{\rm ch}^0}{Ng_{\rm ch}^2g_{\rm d}^2{\mu}_{\rm B}^4}\left({\tilde J}_3^2{\chi}_{\rm d1}^0+{\tilde J}_4^2{\chi}_{\rm d3}^0\right)}\,,\\
\, \nonumber \\ 
&&{\chi}_{\rm d1}={\chi}_{\rm d1}^0\left(1-\frac{6{\tilde J}_3}{Ng_{\rm ch}g_{\rm d}{\mu}_{\rm B}^2}{\chi}_{\rm ch}\right)\,, \\
\, \nonumber \\  
&&{\chi}_{\rm d3}={\chi}_{\rm d3}^0\left(1-\frac{6{\tilde J}_4}{Ng_{\rm ch}g_{\rm d}{\mu}_{\rm B}^2}{\chi}_{\rm ch}\right)\,,
\label{eq:sus3}
\end{eqnarray}
where $N$ is the total number of spins, ${\tilde J}_3\,{=}\,J_3\,{+}\,J^{\prime}_3$ and ${\tilde J}_4\,{=}\,J_4\,{+}\,J^{\prime}_4$, where $J^{\prime}_3$ and $J^{\prime}_4$ are the interlayer exchange interactions between chain spins and dangling spins on Cu(1) and Cu(3), respectively, as shown in Fig.~\ref{fig:exchange}(b). $g_{\rm ch}$ and $g_{{\rm d}}$ are the $g$ factors of the chain spins and dangling spins, respectively. Here we assume that the $g$ factors of spins on the Cu(1) and Cu(3) sites are the same for simplification. ${\chi}_{\rm ch}^0$ is the magnetic susceptibility of the $S\,{=}\,1/2$ alternating antiferromagnetic Heisenberg chain, whose analytical form is given in the literature.\cite{Johnston} ${\chi}_{\rm d1}^0$ and ${\chi}_{\rm d3}^0$ are the magnetic susceptibilities of the dangling spins on the Cu(1) and Cu(3) sites, respectively, which are assumed to obey the Curie-Weiss law ${\chi}_{{\rm d}i}^0\,{=}\,C_i/(T-{\Theta}_i)$. As discussed in the next section, the origin of the Weiss constant ${\Theta}_i$ is considered to be an effective exchange interaction between dangling spins mediated by the chain spins, which causes the ferromagnetic ordering of the dangling spins. 

The total magnetic susceptibility ${\chi}$ is given by ${\chi}\,\,{=}\,\,{\chi}_{\rm ch}\,\,{+}\,\,{\chi}_{\rm d1}\,\,{+}\,\,{\chi}_{\rm d3}$. Fitting eqs.~(\ref{eq:sus1})\,$-$\,(\ref{eq:sus3}) to the experimental susceptibility for $50\,\,{\leq}\,\,T\,\,{\leq}\,\,350$\,K, we obtain $J_1/k_{\rm B}\,{=}\,322$\,K, $J_2/k_{\rm B}\,{=}\,284$\,K, ${\tilde J}_{3,4}/k_{\rm B}\,{=}\,91$\,K, ${\tilde J}_{4,3}/k_{\rm B}\,{=}\,103$\,K, ${\Theta}_{1,3}\,{=}\,12.3$\,K, ${\Theta}_{3,1}\,{=}\,9.6$\,K, $g_{\rm ch}\,{=}\,2.44$ and $g_{\rm d}\,{=}\,2.23$. Since the ${\tilde J}_3$ and ${\tilde J}_4$ interactions are symmetric in eqs.~(\ref{eq:sus1})\,$-$\,(\ref{eq:sus3}), we cannot determine which is larger in the present analysis. The solid line in Fig.~\ref{fig:sus} shows the susceptibility calculated with these parameters. It can be seen that most of the exchange interactions are antiferromagnetic. As expected from the crystal structure, the obtained exchange constants $J_1$ and $J_2$ are about three times as large as ${\tilde J}_3$ and ${\tilde J}_4$ and are similar to each other, i.e., ${\alpha}\,{=}\,J_2/J_1\,{=}\,0.88$. In Cs$_2$Cu$_3$CeF$_{12}$, Cu(2)${-}$F${-}$Cu(2) bond angle $\theta$ for the $J_1$ and $J_2$ interactions is approximately ${\theta}\,{=}\,149^{\circ}$.\cite{Amemiya} The magnitude of the antiferromagnetic exchange interaction increases with increasing the bond angle $\theta$. Hence, the magnitudes of $J_1$ and $J_2$ interactions evaluated by this analysis are valid from the fact that $J/k_{\rm B}\,{=}\,240$\,K and ${\theta}\,{=}\,140^{\circ}$ in Cs$_2$Cu$_3$SnF$_{12}$,\cite{Ono} and $J/k_{\rm B}\,{=}\,390$\,K and ${\theta}\,{=}\,180^{\circ}$ in KCuF$_3$.\cite{Tennant} %The $g$ factor of the dangling spin $g_{\rm d}\,{=}\,2.23$ is consistent with $g_{\rm d}\,{=}\,2.18$ obtained from the $1/3$-plateau shown in Fig.~\ref{fig:MH}. 

The excitation gap of $S\,{=}\,1/2$ alternating Heisenberg antiferromagnetic chain is parameterized well as
\begin{eqnarray}
\label{eq:gap}
{\Delta}(\alpha)\,{=}\,J_1(1\,{-}\,{\alpha})^{3/4}(1\,{+}\,{\alpha})^{1/4},
\end{eqnarray}
for $0\,\,{\leq}\,\,{\alpha}\,\,{\leq}\,\,0.9$.\cite{Barnes} Substituting $J_1/k_{\rm B}\,\,{=}\,\,322$\,\,K and ${\alpha}\,{=}\,0.88$ into eq.~(\ref{eq:gap}), we obtain the gap of the isolated spin chain in Cs$_2$Cu$_3$CeF$_{12}$ as ${\Delta}/k_{\rm B}\,{=}\,77$\,K. In Cs$_2$Cu$_3$CeF$_{12}$, the chain spins are also subjected to the mean field from the dangling spins. Therefore, within the mean field approximation, the critical field $H_{\rm c}$ at which the gap of the spin chain closes is given by  
\begin{eqnarray}
\label{eq:critical}
H_{\rm c}=\frac{1}{g_{\rm ch}{\mu}_{\rm B}}\left\{{\Delta}+\frac{1}{2}\left(J_3+J_3^{\prime}+J_4+J_4^{\prime}\right)\right\}.
\end{eqnarray}
Substituting magnetic parameters obtained by the present analysis into eq.~(\ref{eq:critical}), we estimate the critical field to be $H_{\rm c}\,{=}\,106$\,T, which is twice as large as the highest magnetic field of 54 T in the present experiment.

\begin{figure}[htbp]
\vspace{0.5cm}
\begin{center}
\includegraphics[width=8.5cm, clip]{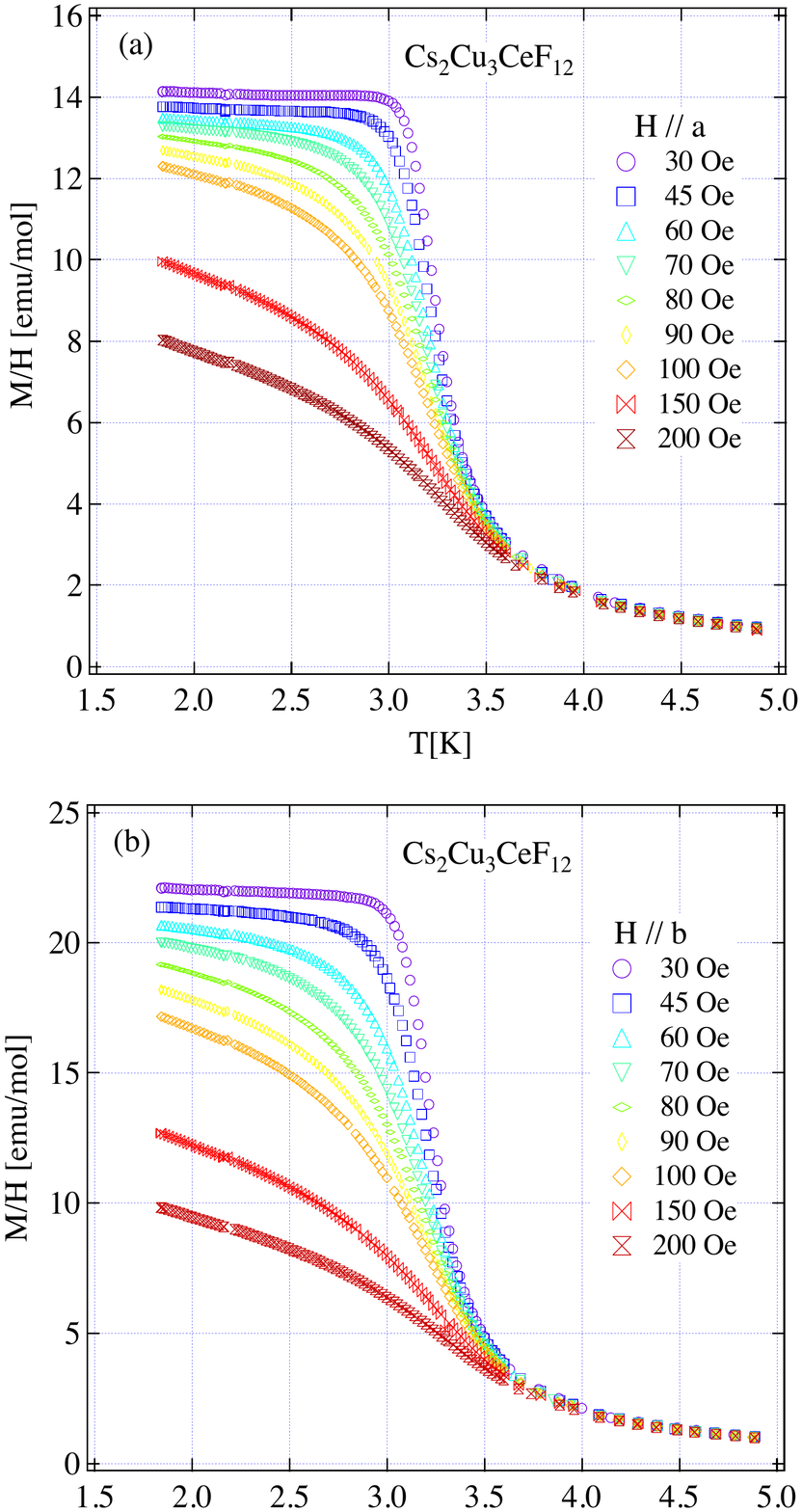}
\end{center}
\caption{(Color online) Low-temperature magnetization divided by magnetic field $M/H$ at various magnetic fields for (a) $H\,{\parallel}\,a$ and (b) $H\,{\parallel}\,b$. }
\label{fig:sus_ab}
\end{figure}

\begin{figure}[htbp]
\vspace{-1.3cm}
\begin{center}
\includegraphics[width=8.5cm, clip]{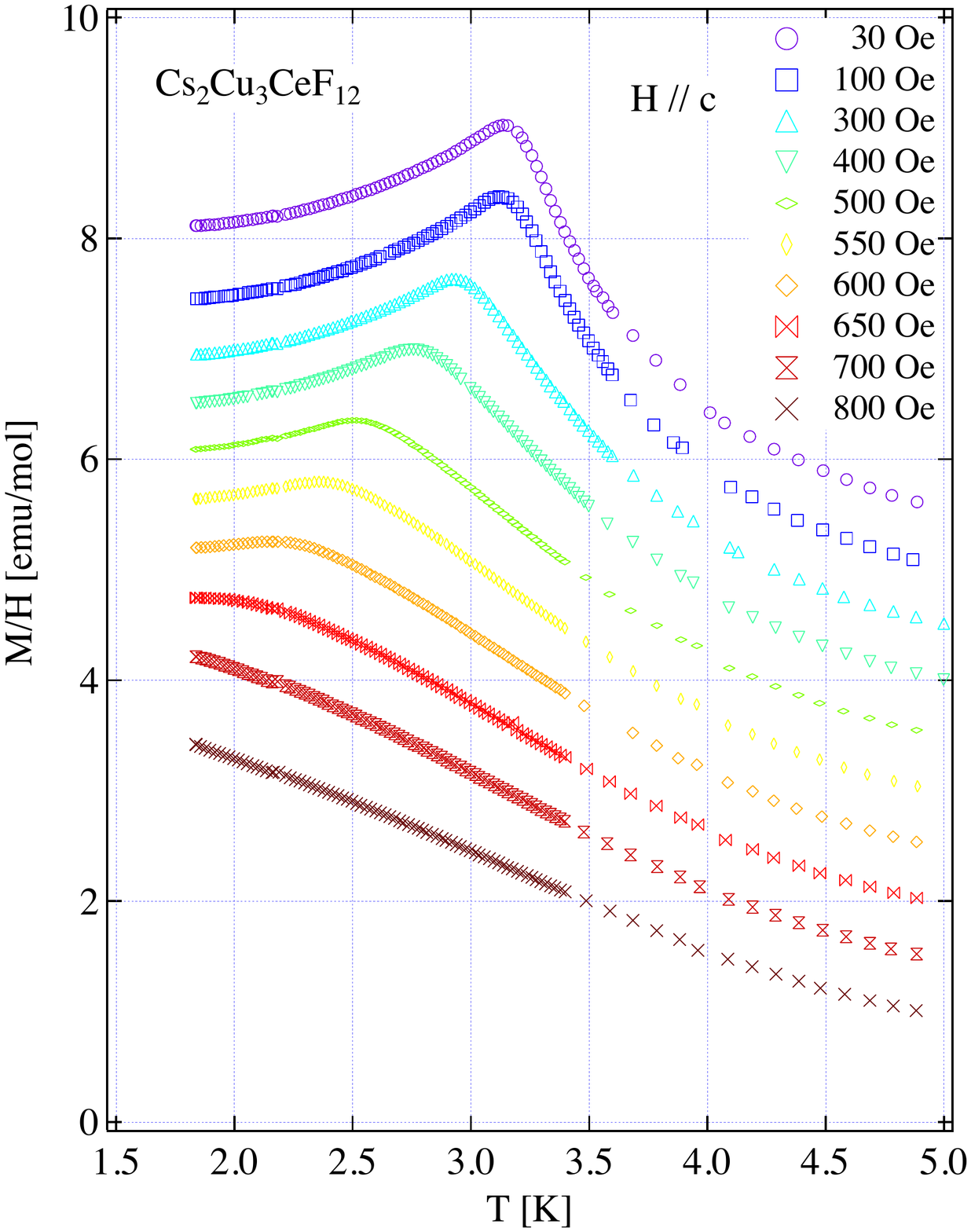}
\end{center}
\vspace{-1.5cm}
\caption{(Color online)  Low-temperature magnetization divided by magnetic field $M/H$ at various magnetic fields for $H\,{\parallel}\,c$. The values of magnetic susceptibility are shifted upward by 0.5 emu/mol with decreasing external field. Arrows denote the transition temperature $T_{\rm C}$.}
\label{fig:sus_c}
\end{figure}

From the sign and magnitude of the Weiss constants ${\Theta}_i$, we see that the effective exchange interaction between dangling spins is ferromagnetic, and that ${\Theta}_i\,{\sim}\,({\bar J}_{\rm d})^2/(k_{\rm B}{\bar J}_{\rm ch})\,{\simeq}\,8$\,K, where ${\bar J}_{\rm ch}\,{=}\,(J_1\,{+}\,J_2)/2$ and ${\bar J}_{\rm d}\,{=}\,(J_3\,{+}\,J_3^{\prime}\,{+}\,J_4\,{+}\,J_4^{\prime})/4$. Because one-half of the dangling spins are paramagnetic, either of the Weiss constants ${\Theta}_i$ should be close to zero. However, both are the same order in magnitude. This failure is considered to arise from the poor accuracy of the present mean field approximation at low temperatures.

\begin{figure}[htbp]
\vspace{-2.5cm}
\begin{center}
\includegraphics[width=8.5cm, clip]{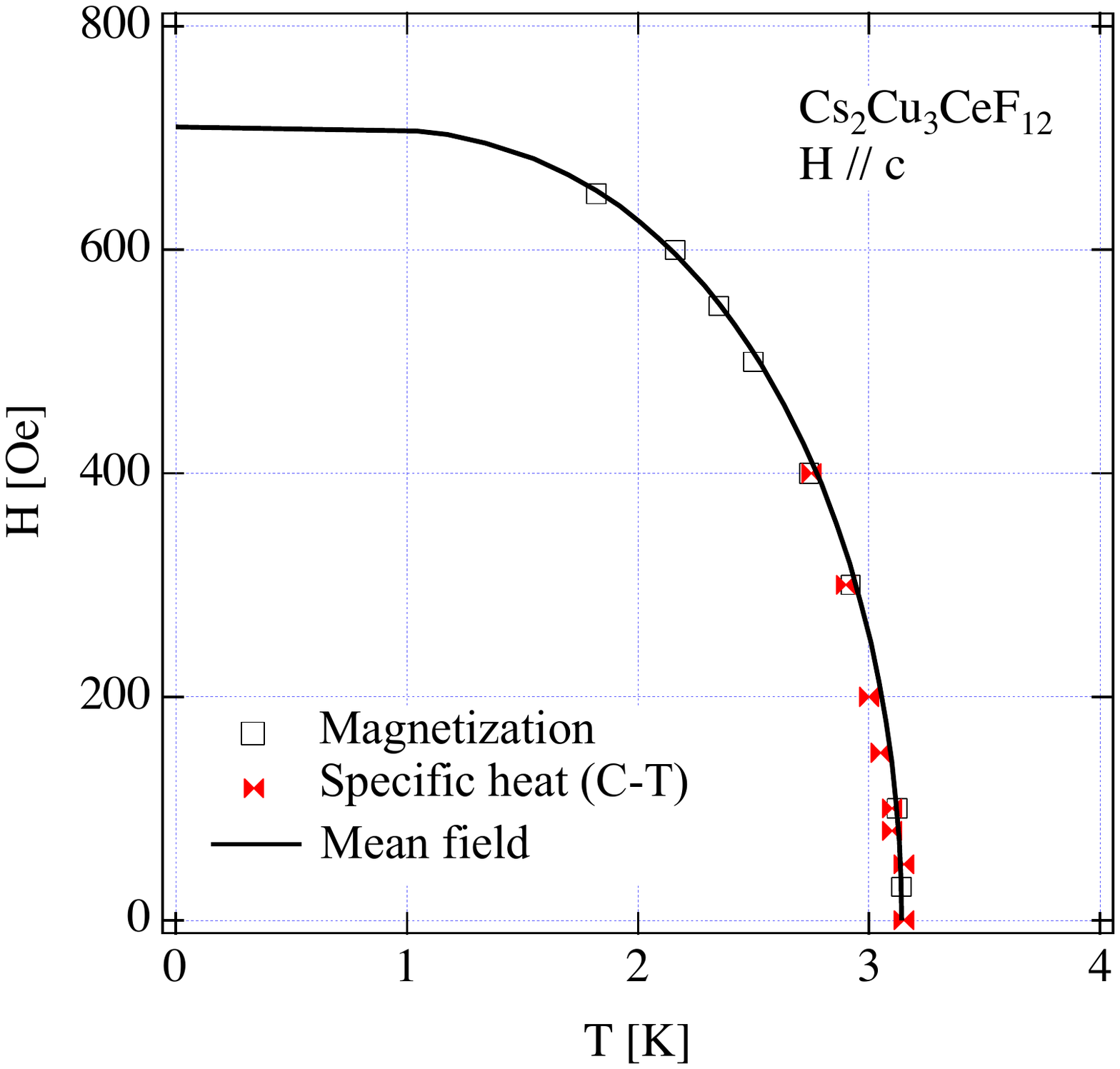}
\end{center}
\vspace{-3cm}
\caption{(Color online) Transition temperature $T_{\rm C}$ obtained from magnetization and specific heat measurements at various magnetic fields for $H\,{\parallel}\,c$. The solid line denotes the phase boundary calculated using eq.~(\ref{eq:t_c}) with $T_{\rm C}(0)\,{=}\,3.14$ K and $H_{\rm c}\,{=}\,710$ Oe.}
\label{fig:phase_c}
\end{figure}

Figures~\ref{fig:sus_ab} and \ref{fig:sus_c} show the temperature dependences of $M/H$ measured at various low magnetic fields for $H\,{\parallel}\,a$, $b$ and $H\,{\parallel}\,c$, respectively. With the exception of the magnitude, values of $M/H$ for $H\,{\parallel}\,a$ and $b$ exhibit similar temperature dependences. With increasing temperature, $M/H$ for $H\,{=}\,30$\,Oe exhibits a sharp increase near $T_{\rm C}\,{=}\,3.14$ K and becomes almost constant. However, with increasing external magnetic field, the increase in $M/H$ around 3 K becomes slow. This indicates that the magnetic ordering undergoes smearing in magnetic fields for $H\,{\parallel}\,a$ and $b$. In contrast to the results for $H\,{\parallel}\,a$ and $b$, $M/H$ for $H\,{\parallel}\,c$ exhibits a cusplike maximum indicative of magnetic ordering. The temperature giving the maximum magnetization decreases with increasing external magnetic field. This means that the magnetic ordering can be defined even in low magnetic fields for $H\,{\parallel}\,c$. We refer to the temperature giving the maximum magnetization as the transition temperature $T_{\rm C}$. The dependence of $T_{\rm C}$ on the magnetic field for $H\,{\parallel}\,c$ is shown in Fig.~\ref{fig:phase_c}.

\begin{figure}[htbp]
\vspace{-2.5cm}
\begin{center}
\includegraphics[width=8.5cm, clip]{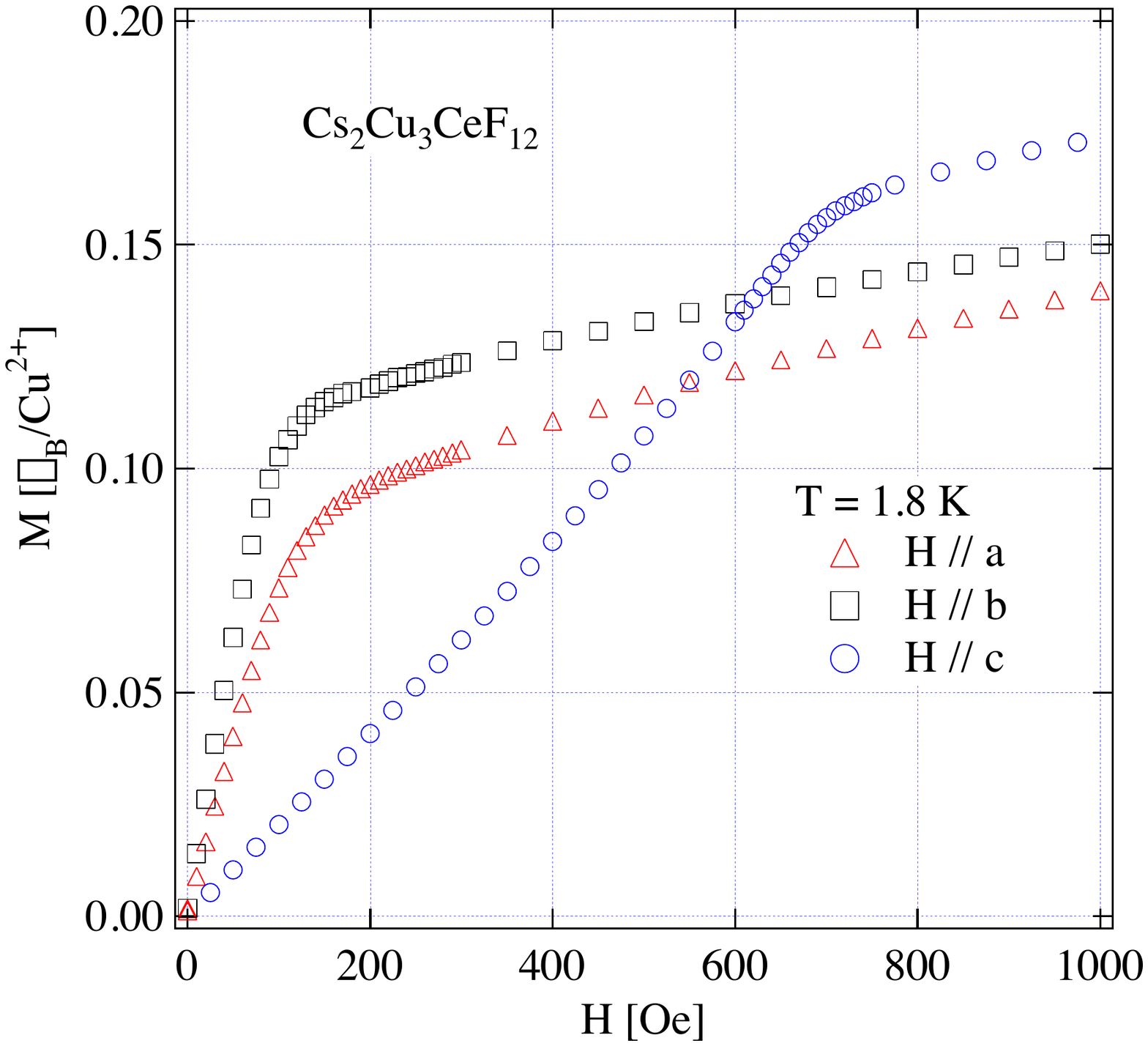}
\end{center}
\vspace{-3cm}
\caption{(Color online) Magnetization curves measured at 1.8 K for external field parallel to the $a$, $b$ and $c$ axes. }
\label{fig:MH2}
\end{figure}

Figure~\ref{fig:MH2} shows magnetization curves measured at 1.8 K for $H\,{\parallel}\,a$, $b$ and $c$. The magnetization for $H\,{\parallel}\,a$ and $b$ exhibits similar dependence on the magnetic field, which is different from that for $H\,{\parallel}\,c$. The magnetization curves for $H\,{\parallel}\,a$ and $b$ display a bending anomaly at $H_{\rm c}\,{\simeq}\,100$ Oe, while for $H\,{\parallel}\,c$, the bending anomaly occurs at $H_{\rm c}\,{\simeq}\,650$ Oe. The magnetization curve for $H\,{\parallel}\,a$ displays hysteresis with a small residual moment of $1\,{\times}\,10^{-2}$\,${\mu}_{\rm B}$/Cu$^{2+}$, as shown in the inset of Fig.~\ref{fig:MH}. The magnetization hysteresis was also observed for $H\,{\parallel}\,b$, while for $H\,{\parallel}\,c$, little hysteresis was observed. Because the sample has a long platelet-like shape with the long axis parallel to the $c$ axis, the internal magnetic fields for $H\,{\parallel}\,a$ and $b$ are reduced by the demagnetizing field, while the internal field for $H\,{\parallel}\,c$ is not reduced. %The internal critical field $H_{\rm c}^{\prime}$ is estimated to be $H_{\rm c}^{\prime}\,{\simeq}\,55$ and 40\, Oe for $H\,{\parallel}\,a$ and $b$, respectively, assuming that the demagnetizing factors for these field directions are $N_a\,{=}\,N_b\,{=}\,2{\pi}$. 

From these magnetization curves, it can be seen that ferromagnetically ordered moments lie in the $ab$ plane at zero magnetic field, and that for $H\,{\parallel}\,c$, the ordered moments cant from the $ab$ plane toward the $c$ axis. The canting angle increases with increasing magnetic field, which causes the slope in the magnetization curve below $H_{\rm c}$ for $H\,{\parallel}\,c$. The transition at $H_{\rm c}$ is interpreted as the transition from a canted ferromagnetic state to the fully polarized state for the ordered dangling DS1 spins. From the temperature and magnetic-field dependences of the magnetization, we can deduce that the effective model for the DS1 spins is approximately described by the ferromagnetic $XXZ$ model with easy-plane anisotropy. 

From the magnetizations above $H_{\rm c}$, it can be seen that the $g$ factor for the dangling DS1 spins is largest when $H\,{\parallel}\,c$. For the elongated CuF$_6$ octahedron, the $g$ factor is largest when the magnetic field is parallel to the elongated axis, while for the compressed octahedron, the $g$ factor is largest when the magnetic field is perpendicular to the compressed axis.\cite{Abragam} The octahedra centered by Cu(1) and Cu(3) are compressed and elongated, respectively, and their compressed and elongated axes are perpendicular to the $c$ axis. Consequently, when $H\,{\parallel}\,c$, the $g$ factor for Cu(1) becomes largest, while for Cu(3), the $g$ factor becomes smallest. Thus, we infer that the spin of Cu(1) is the dangling DS1 spin.

\begin{figure}[htbp]
\vspace{-1cm}
\begin{center}
\includegraphics[width=8.5cm, clip]{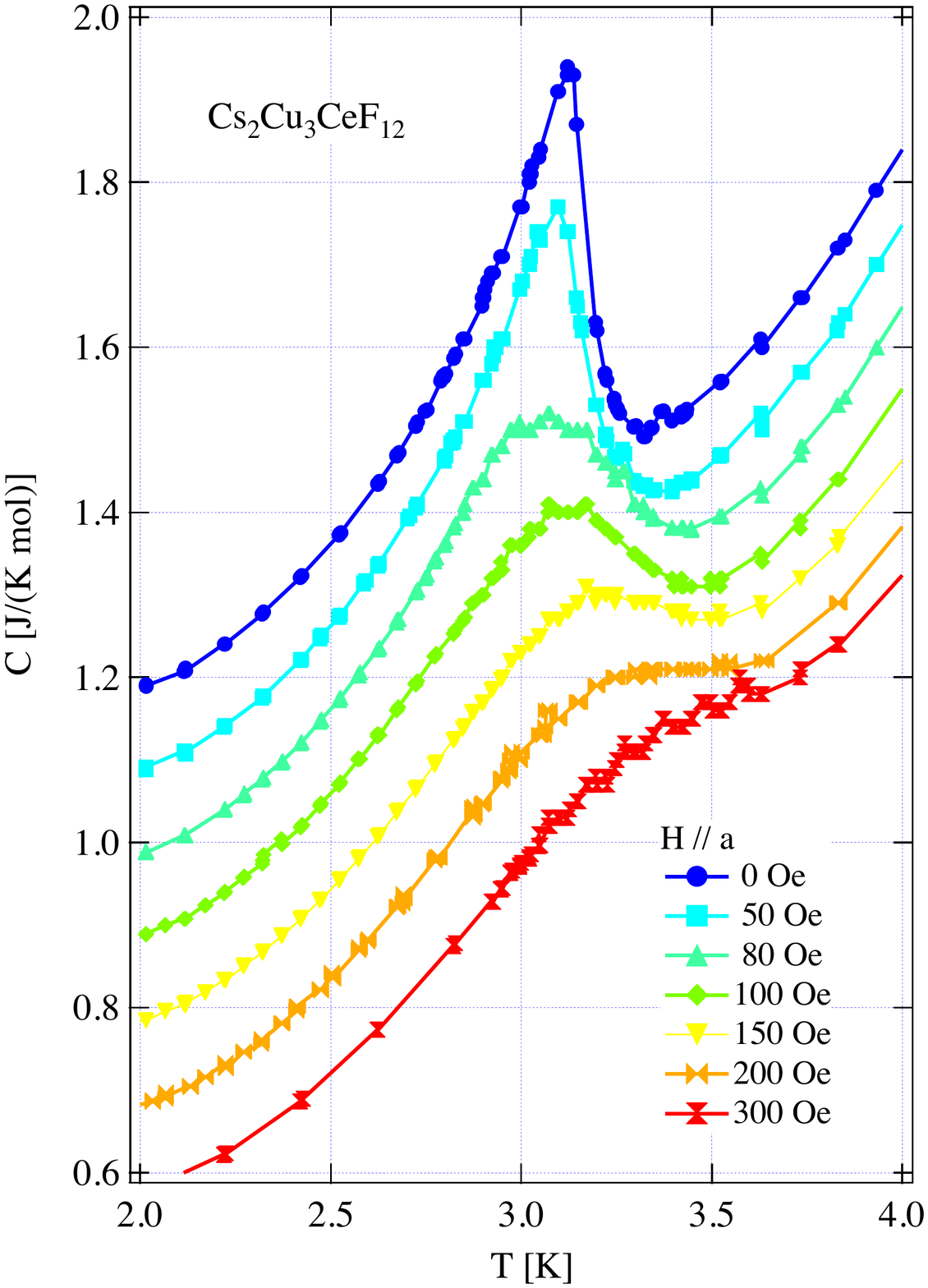}
\end{center}
\vspace{-1.5cm}
\caption{(Color online) Low-temperature specific heat measured at various magnetic fields for $H\,{\parallel}\,a$. The values of specific heat are shifted upward by 0.1 J/(K\,mol) with decreasing external field.}
\label{fig:heat_a1}
\end{figure}

\begin{figure}[htbp]
\vspace{-1cm}
\begin{center}
\includegraphics[width=8.5cm, clip]{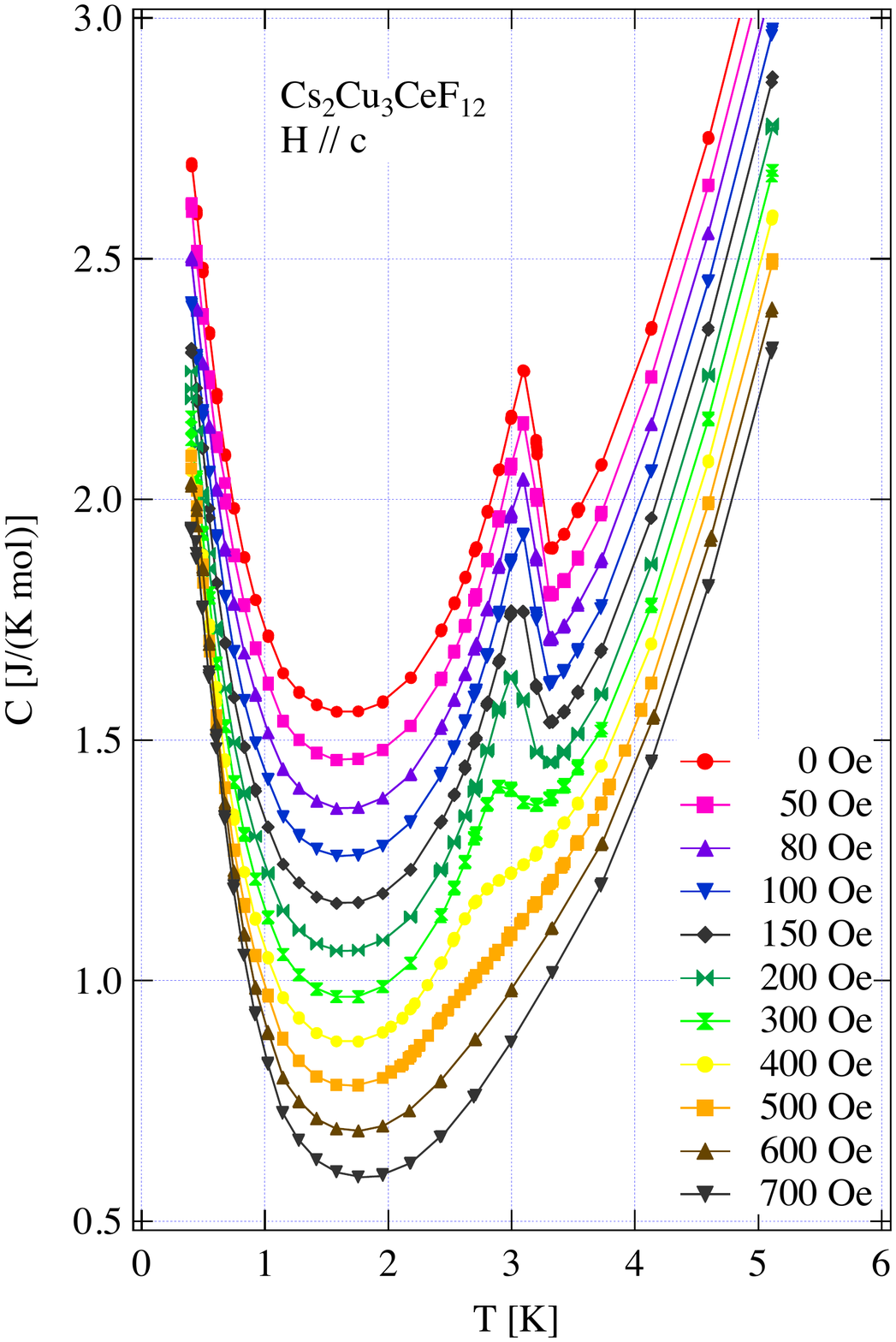}
\end{center}
\vspace{-1cm}
\caption{(Color online) Low-temperature specific heat measured at various magnetic fields for $H\,{\parallel}\,c$. The values of specific heat are shifted upward by 0.1 J/(K\,mol) with decreasing external field.}
\label{fig:heat_c}
\end{figure}

Figures~\ref{fig:heat_a1} and \ref{fig:heat_c} respectively show the low-temperature specific heat for $H\,{\parallel}\,a$ and $c$ measured at various magnetic fields. For $H\,{\parallel}\,a$, the sharp $\lambda$-like anomaly observed at $T_{\rm C}\,{=}\,3.14$ K is rapidly smeared with increasing external magnetic field for $H\,{\geq}\,80$\,Oe. This indicates that the ferromagnetic transition at zero magnetic field changes to a crossover in a finite magnetic field for $H\,{\geq}\,80$\,Oe. When the magnetic field is parallel to the $c$ axis, the $\lambda$-like anomaly indicative of magnetic ordering shifts toward the low-temperature side with increasing magnetic field, as observed in the magnetization measurements. The transition data for $H\,{\parallel}\,c$ are plotted in Fig.~\ref{fig:phase_c} together with those obtained from magnetization measurements. The transition points obtained from the magnetization and specific heat measurements are consistent with each other. The peak height of the $\lambda$-like anomaly becomes smaller with increasing magnetic field. This is because the transverse magnetic moment $m_{\perp}$, which corresponds to the order parameter, decreases with increasing magnetic field.

\section{Discussion} 
The ground state for a spin-$1/2$ Heisenberg antiferromagnet on an extremely spatially anisotropic kagome lattice, which corresponds to the case of $J_1\,{=}\,J_2\,{=}\,J$, $J_3\,{=}\,J_4\,{=}\,J'$ and $J\,{\gg}\,J'$ in Fig.~\ref{fig:exchange}(a), was discussed by Schnyder {\it et al.}\cite{Schnyder} They argued that the ferromagnetic and antiferromagnetic effective exchange interactions ${\cal J}_1$ and ${\cal J}_2$ mediated by the chain spins act between neighboring dangling spins located on both sides of the chain spins and between those located along the $c$ axis, respectively. The magnitudes of ${\cal J}_1$ and ${\cal J}_2$ are of the order of $(J')^2/J$, and ${\cal J}_2/|{\cal J}_1|\,{\simeq}\,0.70$, which lead to a spiral spin structure of chain and dangling spins. This mechanism of effective interactions between dangling spins should be applicable to Cs$_2$Cu$_3$CeF$_{12}$. However, the analytical form of the effective exchange interaction in Cs$_2$Cu$_3$CeF$_{12}$ should be different from that for the uniform chain case discussed by Schnyder {\it et al.}\cite{Schnyder}, because there is a finite gap in the excitation of chain spins, which arises from the exchange alternation along the chain; $J_2/J_1\,{=}\,0.88$. Note that the gap $\Delta$ (${=}\,77$\,K) is much smaller than $J_1$ and $J_2$ (${\simeq}\,300$\,K). Because ${\Theta}_i\,{\sim}\,({\bar J}_{\rm d})^2/(k_{\rm B}{\bar J}_{\rm ch})\,{\sim}\,T_{\rm C}$ as shown in the previous section, we infer that the magnitude of the effective exchange interaction mediated by the chain spins is order of $({\bar J}_{\rm d})^2/(k_{\rm B}{\bar J}_{\rm ch})$, and that the magnetic ordering of the dangling spins is attributed to the effective exchange interaction. Because of the excitation gap in the chain spins, the effective exchange interaction ${\cal J}_2$ mediated by two chain spins should be less effective in Cs$_2$Cu$_3$CeF$_{12}$. 

%From the analysis of the magnetic susceptibility described in the previous section, we can assume that neighboring exchange interactions $J_i$ ($i\,{=}\,1\,{\sim}\,4$), $J_3^{\prime}$ and $J_4^{\prime}$ shown in Fig.~\ref{fig:exchange} are all antiferromagnetic. Because there is no neighboring exchange interaction that couples dangling spins, the 3D ferromagnetic ordering of the dangling spins DS1 cannot be explained by these neighboring exchange interactions. Thus, the 3D ferromagnetic ordering should be attributed to the effective exchange interaction mediated by the chain spins, as discussed by Schnyder {\it et al.}~\cite{Schnyder}.

As illustrated in the exchange network shown in Fig.~\ref{fig:exchange}(b), dangling spins on Cu(1) and Cu(3) couple to chain spins via $J_3$ and $J_3^{\prime}$, and $J_4$ and $J_4^{\prime}$, respectively. The interlayer interaction $J_3^{\prime}$ or $J_4^{\prime}$ is necessary for the 3D ordering of dangling spins. From the analysis of magnetic susceptibility described in the previous section, most of these exchange interactions are considered to be antiferromagnetic. There are three nearest-neighbor effective exchange interactions mediated by one chain spin: ${\cal J}_1(11)$ between the spins on Cu(1) via $J_3$ and $J_3^{\prime}$, ${\cal J}_1(33)$ between the spins on Cu(3) via $J_4$ and $J_4^{\prime}$, and ${\cal J}_1(13)$ between the spins on Cu(1) and Cu(3) via $J_3$ and $J_4$. Because it is considered that the spins on Cu(1) (DS1) undergo 3D ferromagnetic ordering, the effective exchange interaction ${\cal J}_1(11)$ should be ferromagnetic. This means that the $J_3$ and $J_3^{\prime}$ interactions are both antiferromagnetic because ${\cal J}_1(11)$ is proportional to $J_3J_3^{\prime}$. As shown in the previous section, the spins on Cu(3) (DS2) behave as free spins. This indicates that the mean field acting on the DS2 spins vanishes. We infer that the effective exchange interaction ${\cal J}_1(33)$ is antiferromagnetic and cancels out the ferromagnetic effective exchange interaction ${\cal J}_1(13)$. Such a condition can be realized when the exchange interactions $J_4$ and $J_4^{\prime}$ are antiferromagnetic and ferromagnetic, respectively.

The effective exchange interaction ${\cal J}_1(11)$ gives rise to the 3D ordering of the DS1 spins. As shown in the previous section, the effective exchange interaction between the DS1 spins is described by the spin-$1/2$ ferromagnetic $XXZ$ model. Therefore, in a magnetic field parallel to the $c$ axis, the effective Hamiltonian for the DS1 spins should be expressed as 
\begin{eqnarray}
{\cal H}=&-&\sum_{\langle i,j\rangle} \left\{{\cal J}_{ij}^{\perp}(S_i^xS_j^x+S_i^yS_j^y)+{\cal J}_{ij}^{\parallel}S_i^zS_j^z\right\} \nonumber \\
&-&g{\mu}_{\rm B}H\sum_iS_i^z,
\label{eq:model}
\end{eqnarray}
where ${\cal J}_{ij}^{\perp}$ and ${\cal J}_{ij}^{\parallel}$ are the $xy$ and $z$ components of the effective exchange interaction ${\cal J}_1(11)$, respectively, and ${\cal J}_{ij}^{\perp}>{\cal J}_{ij}^{\parallel}>0$. This model given by eq.~(\ref{eq:model}) was discussed by Matsubara and Matsuda,\cite{Matsubara} and was shown to be equivalent to a system of interacting lattice bosons. When ${\cal J}_{ij}^{\perp}>{\cal J}_{ij}^{\parallel}$, a phase transition can occur in a magnetic field. At zero magnetic field the ordered moment ${\bm m}$ lies in the $xy$ plane. With increasing magnetic field, the ordered moment inclines toward the magnetic field. The $xy$ component of the ordered moment ${\bm m}$ gives the order parameter.

In the molecular field approximation,\cite{Matsubara} the relation between the transition temperature $T_{\rm c}(H)$ in a magnetic field $H$ and the reduced magnetization $m^z\,{=}\,2{\langle}S^z{\rangle}$ is given by
\begin{eqnarray}
m^z=\tanh \left\{\frac{T_{\rm C}(0)}{T_{\rm C}(H)}m^z\right\},
\label{eq:t_c}
\end{eqnarray}
where $T_{\rm C}(0)$ is the transition temperature at zero magnetic field. In the ordered state for $H\,{\parallel}\,c$, the magnetization is proportional to $H$, as shown in Fig.~\ref{fig:MH2}. Thus, within the framework of the molecular field approximation, the relation between the magnetic field and $T_{\rm C}(H)$ is given by replacing $m^z$ with $H/H_{\rm c}$ in eq.~(\ref{eq:t_c}), where $H_{\rm c}$ is the saturation field at $T\,{=}\,0$. The solid line in Fig.~\ref{fig:phase_c} indicates the phase boundary calculated using eq.~(\ref{eq:t_c}) with $T_{\rm C}(0)\,{=}\,3.14$ K and $H_{\rm c}\,{=}\,710$ Oe. The experimental and theoretical phase boundaries are in good agreement.

As shown in Fig.~\ref{fig:sus_c}, the magnetization for $H\,{\parallel}\,c$ exhibits a cusplike maximum at the transition temperature $T_{\rm C}(H)$ and decreases below $T_{\rm C}(H)$. This behavior cannot be explained in terms of the molecular field approximation, for which the magnetization is constant below $T_{\rm C}(H)$.\cite{Matsubara} Matsubara and Matsuda \cite{Matsubara} demonstrated that the system of interacting lattice bosons can be mapped onto the  spin model given by eq.~(\ref{eq:model}), and that the magnetic ordering of the spin model is equivalent to the Bose-Einstein condensation (BEC) of the lattice bosons. The magnetization is related to the density of the bosons ${\rho}$ as $m^z\,{=}\,2{\rho}\,{-}\,1$, and the magnetic field corresponds to the chemical potential of the bosons. Because the density of the bosons is larger than $1/2$ at a finite magnetic field, it is convenient to consider the holes instead of the bosons. The density of the holes ${\rho}^{\prime}$ is related to the magnetization as $m^z\,{=}\,1\,{-}\,2{\rho}^{\prime}$. The density of the holes exhibits a cusplike minimum at $T_{\rm C}(H)$ because the magnetization has a cusplike maximum. The cusplike minimum of the density at the BEC transition is characteristic of interacting dilute bosons at a constant chemical potential.\cite{Nikuni,Yamada} The temperature dependence of the density of the holes is explained as follows: above the BEC transition temperature $T_{\rm C}(H)$, the number of thermally excited holes decreases with decreasing temperature. At $T_{\rm C}(H)$, the BEC of holes occurs and the number of condensed holes increases below $T_{\rm C}(H)$ with decreasing temperature. The increase in the number of condensed holes surpasses the decrease in the number of thermally excited holes.\cite{Nikuni,Yamada} For this reason, the density of holes has a minimum at $T_{\rm C}(H)$, which leads to the cusplike maximum of the magnetization. The temperature dependence of the magnetization observed at a low magnetic field for $H\,{\parallel}\,c$ is understandable in terms of the temperature variation of the density of lattice bosons.

\section{Conclusion}
We have reported the results of magnetization and specific heat measurements on Cs$_2$Cu$_3$CeF$_{12}$, which is magnetically described as a spin-$1/2$ spatially anisotropic stacked kagome-lattice antiferromagnet. From the experimental results and their analyses, we found the following. Owing to its strongly spatially anisotropic kagome lattice, Cs$_2$Cu$_3$CeF$_{12}$ can be subdivided into alternating-exchange chains and dangling spins. The dangling spins are further divided into two subsystems DS1 and DS2. The dangling spins in DS1 undergo ferromagnetic ordering at 3.14 K at zero magnetic field, while those in DS2 remain paramagnetic down to 0.35 K. It is considered that he ferromagnetic ordering of the dangling spins in DS1 arises from the effective 3D ferromagnetic exchange interaction mediated by the chain spins. This effective exchange interaction is expressed by the $XXZ$ model with the easy-axis anisotropy, where the $z$ direction is parallel to the crystallographic $c$ axis. For $H\,{\parallel}\,z$, this model is equivalent to the system of interacting lattice bosons.\cite{Matsubara} The magnetic ordering in Cs$_2$Cu$_3$CeF$_{12}$ for $H\,{\parallel}\,c$ can be defined even in magnetic fields. Within the molecular field approximation to the $XXZ$ model, the magnetic field dependence of the ordering temperature of the dangling spins in DS1 observed for $H\,{\parallel}\,c$ can be explained, while the cusplike maximum magnetization at the transition temperature cannot be explained. However, this magnetization behavior is qualitatively understandable in terms of the BEC of interacting lattice bosons.

\begin{acknowledgments}   
This work was supported by a Grant-in-Aid for Scientific Research (A) from the Japan Society for the Promotion of Science (JSPS) and the Global COE Program ``Nanoscience and Quantum Physics'' at Tokyo Institute of Technology funded by the Japanese Ministry of Education, Culture, Sports, Science and Technology. I.U. and T.O. were supported by a JSPS Research Fellowship for Young Scientists and a Grant-in-Aid for Young Scientists (B) from JSPS, respectively. This work was supported in part by a grant from the Mitsubishi Foundation to H.T.
\end{acknowledgments}

%--------------------------------------------------------------------------------------------------------------------


\begin{thebibliography}{99} 
\bibitem{ML} 
G. Misguich and C. Lhuillier, {\it Frustrated Spin Systems}, ed. H. T. Diep (World Science, Singapore, 2005) p. 229.

\bibitem{Zeng1} 
C. Zeng and V. Elser, Phys. Rev. B \textbf{42}, 8436 (1990).
\bibitem{Sachdev} 
S. Sachdev, Phys. Rev. B \textbf{45}, 12377 (1992).
\bibitem{Elstner} 
N. Elstner and A. P. Young, Phys. Rev. B \textbf{50}, 6871 (1994).
\bibitem{Nakamura} 
T. Nakamura and S. Miyashita, Phys. Rev. B \textbf{52}, 9174 (1995).
\bibitem{Lecheminant} 
P. Lecheminant, B. Bernu, C. Lhuillier, L. Pierre, and P. Sindzingre, Phys. Rev. B \textbf{56}, 2521 (1997).
\bibitem{Waldtmann} 
Ch. Waldtmann, H.-U. Everts, B. Bernu, C. Lhuillier, P. Sindzingre, P. Lechminant, and L. Pierre, Eur. Phys. J. B \textbf{2}, 501 (1998).
 
\bibitem{Syromyatnikov}
A. V. Syromyatnikov and S. V. Maleyev, Phys. Rev. B \textbf{66}, 132408 (2002).
\bibitem{Jiang} %Spin gap DMRG, gapless for singlet excitation
H. C. Jiang, Z. Y. Weng, and D. N. Sheng, Phys. Rev. Lett. \textbf{101}, 117203 (2008).
\bibitem{Singh2}%dimer
R. R. P. Singh and D. A. Huse, Phys. Rev. B \textbf{77}, 144415 (2008).
\bibitem{Nikolic} %dimer
P. Nikolic and T. Senthil, Phys. Rev. B \textbf{68}, 214415 (2003).
\bibitem{Budnik} 
R. Budnik and A. Auerbach, Phys. Rev. Lett. \textbf{93}, 187205 (2004).
\bibitem{Yang}%dimer
B.-J. Yang, Y. B. Kim, J. Yu, and K. Park, Phys. Rev. B \textbf{77}, 224424 (2008).
\bibitem{Mambrini} %RVB
M. Mambrini and F. Mila, Eur. Phys. J. B \textbf{17}, 651 (2000).
\bibitem{Hastings}%RVB
M. B. Hastings, Phys. Rev. B \textbf{63}, 014413 (2000).
\bibitem{Ryu} %Critical spin liquid
S. Ryu, O. I. Motrunich, J. Alicea, and M. P. A. Fisher, Phys. Rev. B \textbf{75}, 184406 (2007).

\bibitem{Hermele}
M. Hermele, Y. Ran, P. A. Lee, and X.-G. Wen, Phys. Rev. B \textbf{77}, 224413 (2008).

\bibitem{Mueller} %A$_2$Cu$_3$MF$_{12}$
M. M\"{u}ller and B. G. M\"{u}ller, Z. Anorg. Allg. Chem. \textbf{621}, 993 (1995).
\bibitem{Yamabe}
Y. Yamabe, T. Ono, T. Suto, and H. Tanaka, J. Phys.: Condens. Matter \textbf{19}, 145253 (2007).\bibitem{Morita}
K. Morita, M. Yano, T. Ono, H. Tanaka, K. Fujii, H. Uekusa, Y. Narumi, and K. Kindo, J. Phys. Soc. Jpn. \textbf{77} (2008) 043707.
\bibitem{Ono}
T. Ono, K. Morita, M. Yano, H. Tanaka, K. Fujii, H. Uekusa, Y. Narumi, and K. Kindo, Phys. Rev. B \textbf{79}, 174407 (2009).
\bibitem{Matan} K. Matan, T. Ono, Y. Fukumoto, T. J. Sato, J. Yamaura, M. Yano, K. Morita, and H. Tanaka, Nat. Phys. \textbf{16}, 865 (2010).
\bibitem{Amemiya} T. Amemiya, M. Yano, K. Morita, I. Umegaki, T. Ono, H. Tanaka, K. Fujii, and H. Uekusa, Phys. Rev. B \textbf{80}, 100406R (2009).

\bibitem{Wang} %Theory for J1{\neq}J2
F. Wang, A. Vishwanath, and Y. B. Kim, Phys. Rev. B \textbf{76}, 094421 (2007).
\bibitem{Sindzingre}
P. Sindzingre, arXiv: cond-mat/0707.4264.
\bibitem{Ya}
T. Yavors'kii, W. Apel, and H.-U. Everts, Phys. Rev. B \textbf{76}, 064430 (2007).
\bibitem{Stoudenmire} 
E. M. Stoudenmire and L. Balents, Phys. Rev. B \textbf{77}, 174414 (2008).
\bibitem{Schnyder} 
A. P. Schnyder, O. A. Starykh, and L. Balents, Phys. Rev. B \textbf{78}, 174420 (2008).

\bibitem{Ono1} T. Ono, H. Tanaka, H. Aruga Katori, F. Ishikawa, H. Mitamura, and T. Goto, Phys. Rev. B \textbf{67}, 104431 (2003).
\bibitem{Ono2} T. Ono, H. Tanaka, O. Kolomiyets, H. Mitamura, T. Goto, K. Nakajima, A. Oosawa, Y. Koike, K. Kakurai, J. Klenke, P. Smeibidle, and M. Mei{\ss}ner, J. Phys.: Condens. Matter \textbf{16}, S773 (2004). 
\bibitem{Miyahara} S. Miyahara, K. Ogino, and N. Furukawa, Physica B \textbf{378-380}, 587 (2006).
\bibitem{Fortune}
N. A. Fortune, S. T. Hannahs, Y. Yoshida, T. E. Sherline, T. Ono, H. Tanaka, and Y. Takano, Phys. Rev. Lett. \textbf{102}, 257201 (2009).

%Theory of lattice bosons
\bibitem{Matsubara} T. Matsubara and H. Matsuda, Prog. Theor. Phys. \textbf{16}, 569 (1956).

\bibitem{Johnston} 
D. C. Johnston, R. K. Kremer, M. Troyer, X. Wang, A. Kl\"{u}mper, S. L. Bud'ko, A. F. Panchula, and P. C. Canfield, Phys. Rev. B \textbf{61}, 9558 (2000).
\bibitem{Barnes} T. Barnes, J. Riera, and D. A. Tennant, Phys. Rev. B \textbf{59}, 11384 (1999).

\bibitem{Tennant} 
D. A. Tennant, T. G. Perring, R. A. Cowley, and S. E. Nagler, Phys. Rev. Lett. \textbf{70}, 4003 (1993).

\bibitem{Abragam} A. Abragam and B. Bleaney, {\it Electron Paramagnetic Resonance of Transition Ions}, (Oxford University Press, London, 1970) p. 456.

\bibitem{Nikuni} T. Nikuni, M. Oshikawa, A. Oosawa, and H. Tanaka, Phys. Rev. Lett. \textbf{84}, 5868 (2000). 
\bibitem{Yamada} F. Yamada, T. Ono, H. Tanaka, G. Misguich, M. Oshikawa, and T. Sakakibara, J. Phys. Soc. Jpn. \textbf{77}, 013701 (2008).

\end{thebibliography}
\end{document}